\documentclass[conference]{IEEEtran}
\ifCLASSINFOpdf
\else
\fi

\usepackage{amsbsy,amscd,amsfonts,amsgen,amsmath,amsopn,amssymb,amstext,amsthm,amsxtra}
\usepackage{graphicx}
\usepackage{citesort}
\usepackage{color}
\graphicspath{{./} {img/}}

\IEEEoverridecommandlockouts
\newcounter{cnt01}

\usepackage[margin=0.75in]{geometry}
\newtheorem{CONJ}{Conjecture}
	
\begin{document}
\title{\vspace{0.25in}
Compressed Sensing via Universal Denoising\\ and Approximate Message Passing
\vspace*{-1.7mm}}
\author{\IEEEauthorblockN{
Yanting Ma, Junan Zhu, and Dror Baron}
\IEEEauthorblockA{Department of Electrical and Computer Engineering\\North Carolina State University; Raleigh, NC 27695, USA\\Email: \{yma7, jzhu9, barondror\}@ncsu.edu
\vspace*{-6mm}}
\thanks{This work was supported in part by the National Science Foundation under grant CCF-1217749 and in part by the U.S. Army Research Office under grant W911NF-14-1-0314.}
}
\maketitle
\thispagestyle{empty}
%-------------------
% ABSTRACT
%-------------------
\begin{abstract}
We study compressed sensing (CS) signal reconstruction
problems where an input signal is measured via 
matrix multiplication under additive white Gaussian noise.
Our signals are assumed to be stationary and ergodic,
but the input statistics are unknown; the goal is to 
provide reconstruction algorithms that are universal
to the input statistics.
We present a novel algorithm that combines:
({\em i}) the approximate message passing (AMP) CS reconstruction framework, which converts the matrix channel recovery problem into scalar channel denoising; 
({\em ii}) a universal denoising scheme based on context quantization, 
which partitions the stationary ergodic signal denoising into
independent and identically distributed (i.i.d.) subsequence denoising; and 
({\em iii}) a density estimation approach that approximates the 
probability distribution of an i.i.d. sequence by fitting a 
Gaussian mixture (GM) model. 
In addition to the algorithmic framework, we provide three contributions: 
({\em i}) numerical results showing that state evolution 
holds for non-separable Bayesian sliding-window denoisers;
({\em ii}) a universal denoiser that does not require the input signal to be bounded; and
({\em iii}) we modify the GM learning algorithm, and extend it to an i.i.d. denoiser.
Our universal CS recovery algorithm compares favorably with existing
reconstruction algorithms in terms of both reconstruction quality and
runtime, despite not knowing the input statistics of the stationary
ergodic signal.
\end{abstract}
\IEEEpeerreviewmaketitle

%-----------------------
\section{Introduction}
\label{sec:intro}
%------------------------

%-------------------------
\subsection{Motivation}
\label{subsec:motivation}
%------------------------
Many scientific and engineering problems can be approximated 
as linear systems of the form
\begin{equation}
{\bf y} = {\bf Ax} + {\bf z},
\label{eq:matrix_channel}
\end{equation}
where ${\bf x}\in\mathbb{R}^N$ is the unknown input signal, ${\bf A}\in\mathbb{R}^{M\times N}$ is the matrix that characterizes the linear system, and ${\bf z\in\mathbb{R}}^M$ is measurement noise. The goal is to estimate ${\bf x}$ from the measurements ${\bf y}$ given ${\bf A}$ and statistical information about ${\bf z}$. When $M\ll N$, the setup is known as compressed sensing (CS); by posing a sparsity or compressibility
requirement on the signal, 
it is indeed possible to accurately recover ${\bf x}$ from the ill-posed linear system~\cite{DonohoCS,CandesRUP}. However, we might need $M>N$ when the signal is dense or the noise is substantial.

One popular scheme to solve CS recovery problems is LASSO~\cite{Tibshirani1996} (also known as basis pursuit denoising~\cite{DonohoBP}): $\widehat{{\bf x}}=\arg\min_{{\bf x}\in\mathbb{R}^N} \frac{1}{2}\|{\bf y}-{\bf Ax}\|_2^2+\gamma \|{\bf x}\|_1$, where $\|\cdot\|_p$ denotes the $\ell_p$-norm, and $\gamma$ is a tuning parameter. This approach does not require statistical information about ${\bf x}$ and ${\bf z}$, and can be conveniently solved via standard convex optimization tools or the approximate message passing (AMP) algorithm~\cite{DMM2009}. However, the reconstruction quality is often far from optimal. 

Bayesian CS recovery algorithms based on message passing~\cite{CSBP2010,DMM2010ITW1,RanganGAMP2011ISIT} usually achieve better reconstruction quality, but must know the prior for ${\bf x}$.  
For parametric signals with unknown parameters, one can infer the parameters
and achieve the minimum mean square error (MMSE) in some settings; examples include EM-GM-AMP-MOS~\cite{EMGMTSP}, turboGAMP~\cite{turboGAMP}, adaptive-GAMP~\cite{Kamilov2012}, and AMP-MixD~\cite{MTKB2014ITA}.

Unfortunately, possible uncertainty about the statistics of the signal may make it difficult to select a prior or model class for Bayesian algorithms. Therefore, it would  be desirable to formulate universal algorithms
to estimate ${\bf x}$ that are agnostic to the particular statistics of the signal.

While approaches based on Kolmogorov 
complexity~\cite{DonohoKolmogorov,DonohoKolmogorovCS2006,JalaliMaleki2011,JalaliMalekiRichB2014} are theoretically appealing for universal signal recovery, they are not computable in practice~\cite{Cover06,LiVitanyi2008}.
Several algorithms based on Markov chain Monte Carlo (MCMC)~\cite{BaronFinland2011,BaronDuarteAllerton2011,JZ2014SSP,ZhuBaronDuarte2014_SLAM}
leverage the fact that for stationary ergodic signals,
both the per-symbol empirical entropy and Kolmogorov complexity converge
asymptotically almost surely to the entropy rate of the signal~\cite{Cover06}, and aim to minimize the empirical entropy.
The best existing implementation of the MCMC approach~\cite{ZhuBaronDuarte2014_SLAM} often achieves a mean square error (MSE) 
that is within 3 dB of the MMSE, which resembles a result
by Donoho for universal denoising~\cite{DonohoKolmogorov}.

In this paper, we confine our attention to 
the system model defined in (\ref{eq:matrix_channel}), 
where the input signal ${\bf x}$ is generated by a stationary ergodic source, 
and merge concepts from AMP~\cite{DMM2009}, 
a universal denoising algorithm 
for stationary ergodic signals~\cite{Sivaramakrishnan2008,SW_Context2009},
and Gaussian mixture (GM) learning~\cite{FigueiredoJain2002} for 
density estimation.
The resulting universal CS recovery algorithm, which we call AMP-UD 
(AMP with a universal denoiser), compares 
favorably with existing approaches in terms of reconstruction quality and runtime.

\begin{figure*}[t!]
\setcounter{cnt01}{1}
\vspace*{-6mm}
\center
\includegraphics[width=180mm]{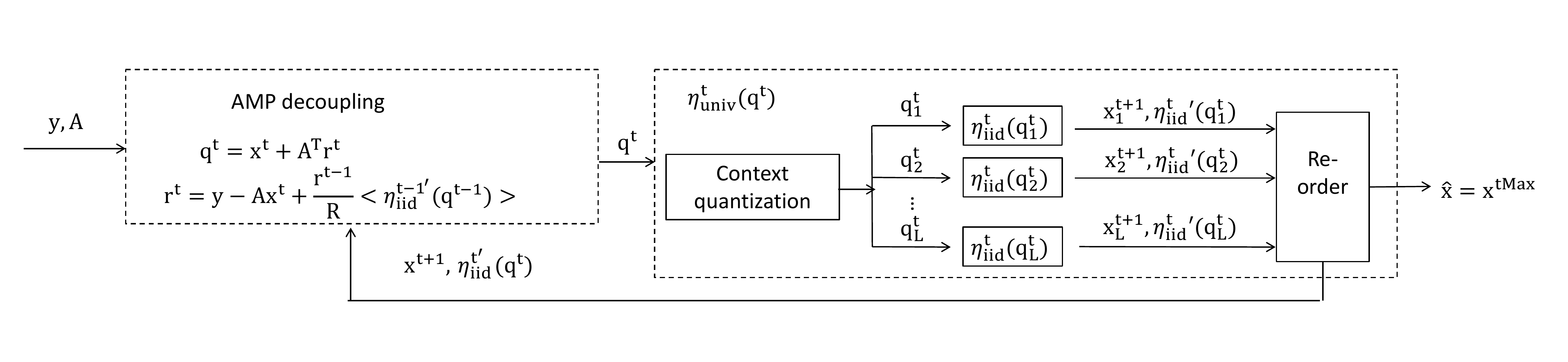}
\vspace*{-10mm}
\caption{Flow chart of AMP-UD. 
AMP decouples the linear inverse problem 
into scalar channel denoising problems. In the $t$-th iteration, the universal denoser 
$\eta^t_{\text{univ}}(\cdot)$ converts stationary ergodic signal denoising into i.i.d. subsequences denoising. Each i.i.d. denoiser $\eta^t_{\text{iid}}(\cdot)$~(\ref{eq:FJ_denoiser}) outputs the denoised subsequence ${\bf x}^{t+1}_l$ and the derivative of the denoiser $\eta_{\text{iid}}^{t'}(\cdot)$~(\ref{eq:denoiser_FJ_deriv}). The algorithm stops when the iteration index $t$ reaches the predefined maximum tMax, and outputs 
$\widehat{\bf x}^{\text{tMax}}$ as the CS recovery result.}
\label{fig:FlowChart}
\vspace*{-5mm}
\setcounter{figure}{\value{cnt01}}
\end{figure*}

%----------------------------
\subsection{Related work and main results}
\label{subsec:contribution}
%----------------------------
{\bf Approximate message passing:} 
AMP is an iterative algorithm that solves a linear inverse problem by successively converting the matrix channel problem into scalar channel denoising
problems with additive white Gaussian noise (AWGN). AMP has received considerable attention because of its fast convergence and the state evolution (SE) formalism~\cite{DMM2009,Bayati2011}, which offers a precise
characterization of the AWGN denoising problem in each iteration. 
AMP with separable denoisers has been rigorously proved to obey SE~\cite{Bayati2011}. However, for non-i.i.d. signals we may want to explore non-separable denoisers. 
Donoho et al.~\cite{Donoho2013} provide numerical results demonstrating 
that SE accurately predicts the phase transition of AMP when some well-behaved non-separable minimax denoisers are applied, and conjecture that SE holds for AMP with a broader class of denoisers. 
A compressive imaging algorithm that applies non-separable image denoisers within AMP appears in Tan et al.~\cite{Tan_CompressiveImage2014}.
A potential challenge of implementing AMP is to obtain the
Onsager correction term~\cite{DMM2009}, 
which involves the calculation of the derivative of a denoiser.
Metzler et al.~\cite{Metzler2014} leverage a Monte Carlo technique to approximate the derivative of a denoiser when an explicit input-output relation of the denoiser is not available, and provide numerical results showing that SE holds for AMP with their approximation.

Despite the encouraging results for using non-separable denoisers within AMP,
a rigorous proof that SE holds for non-separable denoisers has yet to appear.
Consequently, new evidence showing that AMP obeys SE 
may increase the community's confidence about using non-separable 
denoisers within AMP. Our first contribution is that we provide numerical results showing that SE holds for non-separable Bayesian sliding-window denoisers. 

{\bf Universal denoising:} 
Our proposed denoiser is inspired by an approach based on context 
quantization~\cite{SW_Context2009}, where a universal denoiser for a 
stationary ergodic signal involves multiple separable denoisers for 
conditionally independent subsequences. Sivaramakrishnan and 
Weissman~\cite{SW_Context2009} have shown 
that their universal denoiser based on context quantization 
can achieve the MMSE for stationary ergodic signals with bounded components.

The boundedness condition of Sivaramakrishnan and 
Weissman~\cite{SW_Context2009}
is due to their density estimation approach, 
in which the empirical distribution function
is obtained by quantizing the bounded range of the signal.
Such boundedness conditions may be undesirable in certain applications. 
We overcome this limitation by replacing their density estimation approach 
with GM model learning.  Our second contribution is a universal denoiser that does not require the input signal to be bounded; we conjecture that our universal denoiser achieves the MMSE under some technical conditions.

{\bf Fitting Gaussian mixture models:}
Figueiredo and Jain~\cite{FigueiredoJain2002} propose an algorithm that fits 
a given data sequence with a GM model. The algorithm employs a cost function
that resembles the minimum message length (MML) criterion, and the parameters
are learned using a modified expectation-maximization (EM) algorithm.

Our GM fitting problem involves a sequence of observations 
${\bf q}$ corrupted by
AWGN, and we estimate the probability density function (pdf)
of ${\bf x}$ based on the GM model for ${\bf q}$. 
Note that a GM convolved with Gaussian noise is still a GM in which the
variance of each component is increased by the noise variance. 
Therefore, we send ${\bf q}$ to the mixture model learning algorithm, 
and subtract the noise variance from each component of the estimated 
{pdf} $\widehat{p}(q)$ to obtain $\widehat{p}(x)$. Once 
$\widehat{p}(x)$ is available, we denoise the subsequence by computing 
the conditional expectation of each entry of the subsequence of ${\bf x}$ 
based on the estimated prior $\widehat{p}(x)$.\footnote{We remind the
reader that MMSE-optimal estimators rely on conditional expectation.} Our third contribution is that we modify the GM learning algorithm, and extend it to an i.i.d. denoiser.

A flow chart of AMP-UD, which employs the AMP framework, along with our modified universal denoiser ($\eta_{\text{univ}}$) and the GM-based i.i.d. denoiser ($\eta_{\text{i.i.d.}}$), is shown in Fig.~\ref{fig:FlowChart}. 
Based on the numerical evidence that SE holds for AMP with the Bayesian sliding-window denoiser and the conjecture that our universal denoiser can achieve the MMSE, we further conjecture that AMP-UD achieves the MMSE under some technical conditions.
The details of AMP-UD are developed in Sections~\ref{sec:AMP}--\ref{sec:proposed_algo}.

The remainder of the paper is arranged as follows. 
In Section~\ref{sec:AMP}, we review AMP and provide new numerical evidence 
that AMP obeys SE with non-separable denoisers. 
In Section~\ref{sec:SW}, we extend the universal denoiser based on
context quantization  to overcome the boundedness condition.
Section~\ref{sec:FJ} modifies the GM fitting algorithm, and extends it to an i.i.d. denoiser. Our proposed AMP-UD algorithm 
is described in detail in Section~\ref{sec:proposed_algo}. Numerical
results are shown in Section~\ref{sec:numerical}, and we conclude the paper in Section~\ref{sec:conclusion}.

%----------------------------
\section{Approximate message passing for sliding-window denoisers}
\label{sec:AMP}
%----------------------------

%---------------------------
\subsection{Review of AMP}
\label{subsec:AMP_review}
%--------------------------
Consider a linear inverse problem (\ref{eq:matrix_channel}), where the empirical pdf of ${\bf x}$ follows $p_X(x)$, the measurement matrix~${\bf A}$ has i.i.d. Gaussian entries with unit-norm columns on average, and ${\bf z}\sim\mathcal{N}({\bf z};{\bf 0},\sigma_z^2I)$, where $\mathcal{N}({\bf u};{\boldsymbol\mu},{\boldsymbol\Sigma})=\frac{1}{\sqrt{(2\pi)^n|{\boldsymbol\Sigma}|}}\exp(-\frac{1}{2}({\bf u}-{\boldsymbol\mu})^T{\boldsymbol\Sigma}^{-1}({\bf u}-{\boldsymbol\mu}))$ denotes a multivariate Gaussian pdf of a random vector ${\bf u}\in\mathbb{R}^n$, $(\cdot)^T$ denotes the transpose, and $I$ is the identity matrix.   

Starting with ${\bf x}^0=0$, the AMP algorithm~\cite{DMM2009} proceeds iteratively according to
\begin{align}
{\bf x}^{t+1}&=\eta^t({\bf A}^T{\bf r}^t+{\bf x}^t)\label{eq:AMPiter1},\\
{\bf r}^t&={\bf y}-{\bf Ax}^t+\frac{1}{R}{\bf r}^{t-1}
\langle\eta^{t-1'}({\bf A}^T{\bf r}^{t-1}+{\bf x}^{t-1})\rangle\label{eq:AMPiter2},
\end{align}
where~$R=M/N$ represents the measurement rate, $t$ represents the iteration index, $\eta^t(\cdot)$ is a denoising function, and~$\langle{\bf u}\rangle=\frac{1}{N}\sum_{i=1}^N u_i$
for some vector~${\bf u}\in\mathbb{R}^N$.
The denoising function~$\eta^t(\cdot)$ is separable in the original derivation of AMP~\cite{DMM2009,Montanari2012,Bayati2011}. That is, $\eta^t({\bf u})=(\eta^t(u_1),\eta^t(u_2),...,\eta^t(u_N))$ and $\eta_t'({\bf u})=(\eta^{t'}(u_1),\eta^{t'}(u_2),...,\eta^{t'}(u_N))$, where $\eta^{t'}(\cdot)$ denotes the derivative of $\eta^t(\cdot)$. 
A useful property of AMP is that at each iteration, 
the vector~${\bf A}^T{\bf r}^t+{\bf x}^t\in\mathbb{R}^N$ 
in (\ref{eq:AMPiter1}) is equivalent to 
the input signal ${\bf x}$ corrupted by AWGN. The variance of the Gaussian noise~$(\sigma^t)^2$ evolves following SE in the limit of large systems ($N\rightarrow\infty, M/N\rightarrow R$):
\begin{equation}
(\sigma^{t+1})^2=\sigma^2_z+\frac{1}{R}\mathbb{E}_{X,W}\left[\left( \eta^t\left( X+\sigma^{t}W \right)-X \right)^2\right]\label{eq:SE},
\end{equation}
where $W\sim\mathcal{N}(w;0,1)$, $X\sim p_X(x)$, and $(\sigma^0)^2=\sigma_z^2+\frac{1}{R}\mathbb{E}[X^2]$.
Formal statements for SE appear 
in~\cite{Bayati2011,Montanari2012}. 
Additionally, it is convenient to use the following unbiased
estimator for $(\sigma^t)^2$\cite{Montanari2012}:
\begin{equation}
(\widehat{\sigma}^t)^2 = \frac{1}{M}\|{\bf r}^t\|_2^2.
\label{eq:AMP_noise_est}
\end{equation}

%-----------------------------------------------------------------
\subsection{State evolution for Bayesian sliding-window denoisers}
\label{subsec:SE}
%-----------------------------------------------------------------

It has been conjectured by Donoho et al.~\cite{Donoho2013} that 
AMP with a wide range of denoisers obeys SE, including many
non-separable denoisers.
We now provide new evidence to support this conjecture by constructing 
non-separable Bayesian denoisers within the sliding-window denoising scheme for two Markov signal models, and showing that SE accurately predicts the performance of AMP with this class of denoisers for large signal dimension $N$.
Our rationale for examining the SE performance of sliding-window
denoisers is that the context quantization approach of
Sivaramakrishnan and Weissman~\cite{SW_Context2009} resembles
a sliding-window denoiser. 
 
The mathematical model for an AWGN channel denoising problem is defined as
\begin{equation}
{\bf q}={\bf x}+{\bf v},
\label{eq:denoising}
\end{equation}
where ${\bf x}\in\mathbb{R}^N$ is the input signal, ${\bf v}\in\mathbb{R}^N$ is AWGN that follows the distribution $\mathcal{N}({\bf v};{\bf 0},\sigma_v^2I)$, and ${\bf q}\in\mathbb{R}^N$ is a sequence of noisy observations. 

In a separable
denoiser, $x_j$ is estimated only from its noisy observation $q_j$. The separable Bayesian denoiser that minimizes the MSE is the point-wise conditional expectation,
\begin{equation}
\widehat{x}_j=\mathbb{E}[x_j|q_j]=\int x_jp(x_j|q_j)\text{d}x_j,\label{eq:Bayes_scalar}
\end{equation}
where Bayes' rule yields $p(x_j|q_j)=\frac{p(y_j|x_j)p(x_j)}{p(q_j)}$. If entries of the input signal ${\bf x}$ are independent and $x_j$ is drawn from $p(x_j)$, then (\ref{eq:Bayes_scalar}) achieves the MMSE.

When there are statistical dependencies among the entries of ${\bf x}$, a sliding-window denoising scheme can be applied to improve performance. We consider two Markov sources as examples that contain statistical
dependencies, and highlight that our true motivation is the richer 
class of stationary ergodic sources.

{\bf Example source~1:} Consider a two-state Markov state machine that
contains states $s_0$ (zero state) and $s_1$ (nonzero state). 
The transition probabilities are $p_{10}=p(s_0|s_1)$ and 
$p_{01}=p(s_1|s_0)$. In the steady state, the marginal 
probability of state $s_1$ is $p(s_1)=\frac{p_{01}}{p_{01}+p_{10}}$.
We call our first example source Markov-constant (MConst for short);
it is generated by the two-state Markov machine with $p_{01}=\frac{3}{970}$ 
and $p_{10}=0.10$, 
and in the nonzero state the signal value is the constant 1.
These state transition parameters yield 3\% nonzero 
entries in a MConst signal on average. 

{\bf Example source~2:} Our second example is
a four-state Markov switching signal (M4 for short) that follows 
the pattern $+1,+1,-1,-1,+1,+1,-1,-1...$ with 3\% error probability in state transitions, resulting in the signal switching from $-1$ to $+1$ or vice versa either too early or too late; the four states $s_1=[-1\ -1]$, $s_2=[-1\ +1]$, $s_3=[+1\ -1]$, 
and $s_4=[+1\ +1]$ have equal marginal probabilities $0.25$ in 
the steady state.

{\bf Bayesian sliding-window denoiser:}
Denote a block $(q_s,q_{s+1},...,q_{t})$ of a sequence ${\bf q}$ by ${\bf q}_{s}^{t}$ for $s<t$. The $(2k+1)$-Bayesian sliding-window denoiser $\eta_\text{MConst}$ for the MConst signal is defined as
\begin{equation}
\eta_{\text{MConst},j}({\bf q}_{j-k}^{j+k})=\mathbb{E}[x_j|{\bf q}_{j-k}^{j+k}]=\frac{p(x_j=1,{\bf q}_{j-k}^{j+k})}{p({\bf q}_{j-k}^{j+k})},\label{eq:Bayes_MConst}
\end{equation}
and the MSE of $\eta_\text{MConst}$ can be shown to be
\begin{align}
&\text{MSE}^{\text{MConst}}=\mathbb{E}\left[\left(x_j-\eta_{\text{MConst},j}({\bf q}_{j-k}^{j+k})\right)^2\right]\nonumber\\
&= \int\frac{p(x_j=0,{\bf q}_{j-k}^{j+k})p(x_j=1,{\bf q}_{j-k}^{j+k})}{p({\bf q}_{j-k}^{j+k})} \text{d}{\bf q}_{j-k}^{j+k},\label{eq:Bayes_MConst_MSE}
\end{align}
where
\begin{align}
p(x_j=s,{\bf q}_{j-k}^{j+k})&=\sum_{{\bf x}_{j-k}^{j+k}\setminus x_j}\mathcal{N}({\bf q}_{j-k}^{j+k};{\bf x}_{j-k}^{j+k},\sigma_v^2I)\left.p({\bf x}_{j-k}^{j+k})\right\vert_{x_j=s},\label{eq:Bayes_MConst_1}
\end{align}
${\bf x}_{j-k}^{j+k}\setminus x_j$ denotes the sequence ${\bf x}_{j-k}^{j+k}$ not including the middle symbol $x_j$,
\begin{align}
p({\bf q}_{j-k}^{j+k})&=\sum_{{\bf x}_{j-k}^{j+k}}\mathcal{N}({\bf q}_{j-k}^{j+k};{\bf x}_{j-k}^{j+k},\sigma_v^2I)p({\bf x}_{j-k}^{j+k}),\label{eq:Bayes_MConst_2}\\
p({\bf x}_{j-k}^{j+k})&=p(x_{j-k})\prod_{t=-k}^{k-1}p(x_{j+t+1}|x_{j+t})\label{eq:Bayes_MConst_3}.
\end{align}

To obtain the Onsager correction term 
$\frac{1}{R}{\bf r}^{t-1}$
$\langle\eta^{{t-1}'}({\bf A}^T{\bf r}^{t-1}+{\bf x}^{t-1})\rangle$ in (\ref{eq:AMPiter2}), we need to calculate the derivative of $\eta_{\text{MConst},j}$. It can be shown that
\begin{equation}
\frac{\partial}{\partial q_j}\eta_{\text{MConst},j}({\bf q}_{j-k}^{j+k})=\frac{p(x_j=0,{\bf q}_{j-k}^{j+k})p(x_j=1,{\bf q}_{j-k}^{j+k})}{\left(\sigma_v p({\bf q}_{j-k}^{j+k})\right)^2}.\label{eq:Bayes_MConst_deriv}
\end{equation}

Similarly, the $(2k+1)$-Bayesian sliding-window denoiser $\eta_\text{M4}$ for the M4 signal is defined as
\begin{align}
\eta_{\text{M4},j}({\bf q}_{j-k}^{j+k})&=\mathbb{E}[x_j|{\bf q}_{j-k}^{j+k}]\nonumber\\
&=\frac{p(x_j=1,{\bf q}_{j-k}^{j+k})-p(x_j=-1,{\bf q}_{j-k}^{j+k})}{p({\bf q}_{j-k}^{j+k})},\label{eq:Bayes_M4}
\end{align}
where $p(x_j=s,{\bf q}_{j-k}^{j+k})$ for $s\in\lbrace -1,1 \rbrace$ is defined in (\ref{eq:Bayes_MConst_1}), $p({\bf q}_{j-k}^{j+k})$ is defined in (\ref{eq:Bayes_MConst_2}), and (\ref{eq:Bayes_MConst_3}) becomes 
\begin{equation*}
p({\bf x}_{j-k}^{j+k})=p(x_{j-k+1},x_{j-k})\prod_{t=-k}^{k-2}p(x_{j+t+2}|x_{j+t+1},x_{j+t}).
\end{equation*}
It can be shown that
\begin{align}
&\text{MSE}^{\text{M4}}=\mathbb{E}\left[\left(x_j-\eta_{\text{M4},j}({\bf q}_{j-k}^{j+k})\right)^2\right]\nonumber\\
&= \int\frac{4p(x_j=0,{\bf q}_{j-k}^{j+k})p(x_j=1,{\bf q}_{j-k}^{j+k})}{p({\bf q}_{j-k}^{j+k})} \text{d}{\bf q}_{j-k}^{j+k},\label{eq:Bayes_M4_MSE}\\
&\frac{\partial}{\partial q_j}\eta_{\text{M4},j}({\bf q}_{j-k}^{j+k})=\frac{4p(x_j=0,{\bf q}_{j-k}^{j+k})p(x_j=1,{\bf q}_{j-k}^{j+k})}{\left(\sigma_z p({\bf q}_{j-k}^{j+k})\right)^2}.\label{eq:Bayes_M4_deriv}
\end{align}

\begin{figure}[t!]
\center
\includegraphics[width=85mm]{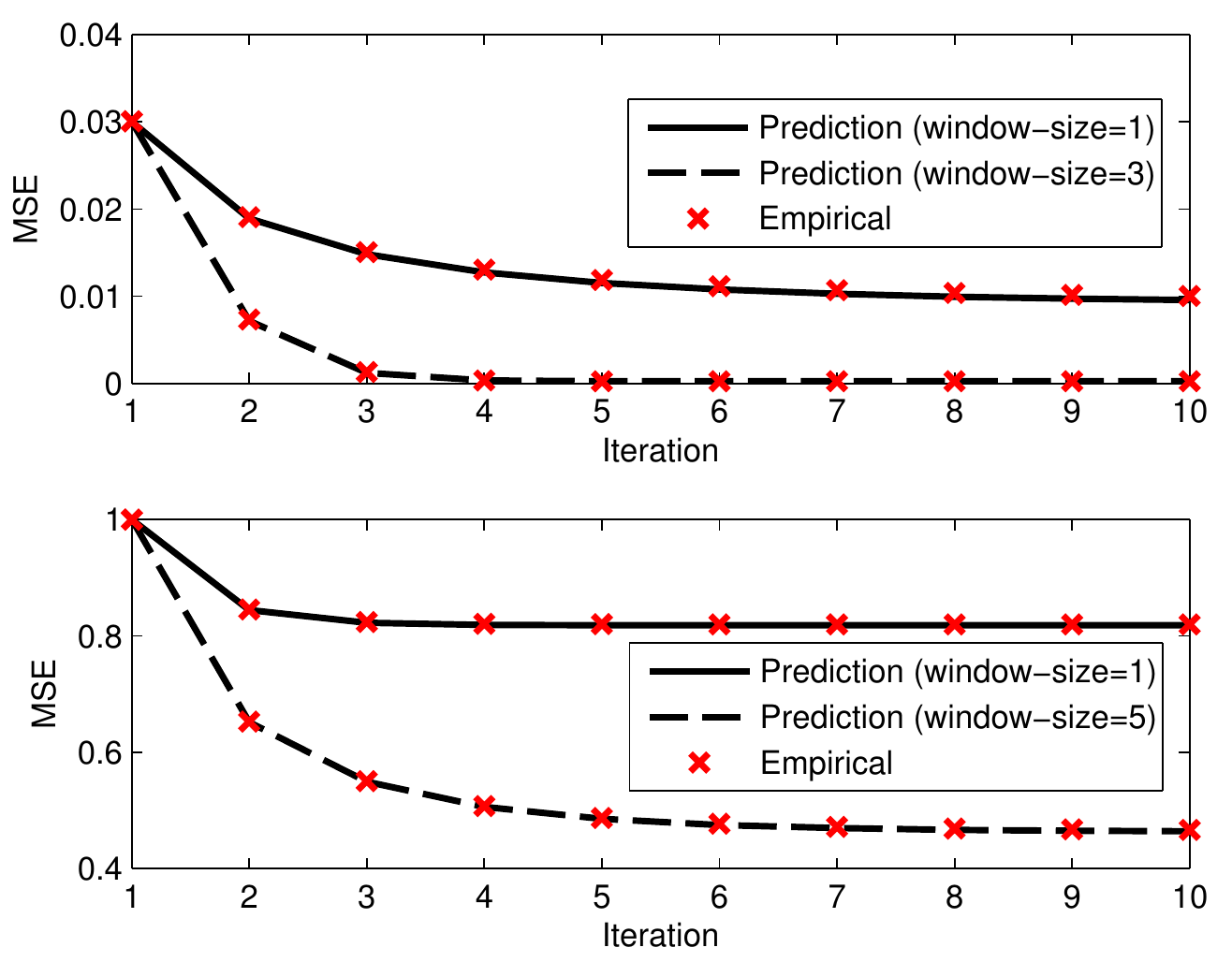}
\caption{Top: Numerical verification of SE for AMP with $\eta_{\text{MConst}}$ (\ref{eq:Bayes_MConst}) when the input is a MConst signal. ($N=20,000, R=0.2, \text{SNR}=5 \text{ dB}$.) Bottom: Numerical verification of SE for AMP with $\eta_{\text{M4}}$ (\ref{eq:Bayes_M4}) when the input is a M4 signal. ($N=20,000, R=0.2, \text{SNR}=10 \text{ dB}$.)}
\label{fig:SE_verify}
\vspace*{-5mm}
\end{figure}

If AMP with $\eta_\text{MConst}$ or $\eta_\text{M4}$ obeys SE, 
then the noise variance $(\sigma^t)^2$ at each iteration should evolve 
according to (\ref{eq:SE}). As a consequence, 
the reconstruction error at iteration $t$ can be predicted by evaluating (\ref{eq:Bayes_MConst_MSE}) or (\ref{eq:Bayes_M4_MSE}) with $\sigma_v^2$ in (\ref{eq:Bayes_MConst_1}) and (\ref{eq:Bayes_MConst_2}) being replaced by $(\sigma^t)^2$.

%---------------------------------------
\subsection{Numerical evidence}
\label{subsec:evidence_noniid}
%---------------------------------------

We apply $\eta_{\text{MConst}}$ (\ref{eq:Bayes_MConst}) within AMP 
for MConst signals, and $\eta_{\text{M4}}$ (\ref{eq:Bayes_M4}) within
AMP for M4 signals. The window size $2k+1$ is chosen to be 1 or 3 for  $\eta_{\text{MConst}}$, and 1 or 5 for  $\eta_{\text{M4}}$. Note that when the window size is 1,  $\eta_{\text{MConst}}$ and  $\eta_{\text{M4}}$ become separable denoisers. The MSE predicted by SE is compared to the empirical MSE at each iteration where the input signal to noise ratio ($\text{SNR}=10\log_{10}[(N\mathbb{E}[x^2])/(M\sigma_z^2)]$) is 5 dB or 10 dB.
It is shown in Fig.~\ref{fig:SE_verify} 
for AMP with $\eta_{\text{MConst}}$ and $\eta_{\text{M4}}$
that the markers representing the empirical MSE track the lines predicted by SE, 
and that side-information from neighboring entries helps 
improve the MSE performance.

\begin{figure}[t!]
\center
\includegraphics[width=85mm]{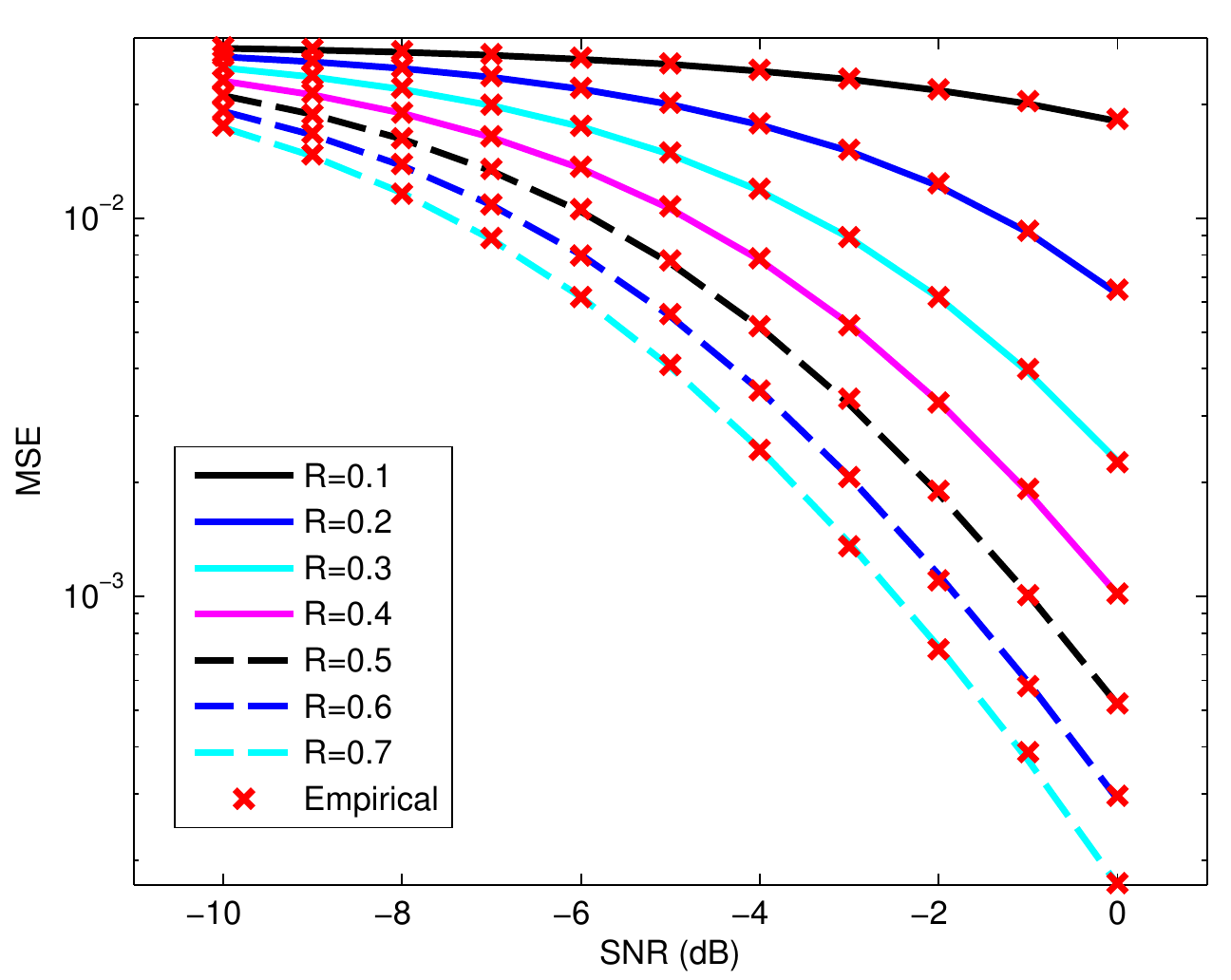}
\caption{ Numerical verification of SE for AMP with $\eta_{\text{MConst}}$ (\ref{eq:Bayes_MConst}) at various measurement rates ($R$) and noise levels (SNR). The lines are predicted by SE, whereas the markers represent the empirical MSE. ($N=20,000$.)}
\label{fig:SE_verify1}
\vspace*{-5mm}
\end{figure}

Results for AMP with $\eta_{\text{MConst}}$ at various measurement rates and noise levels are shown in Fig.~\ref{fig:SE_verify1}. The markers that represent the empirical MSE lie on the lines predicted by SE, which further verifies the correctness of SE.

Our SE results for the two Markov signals increase
our confidence that applying non-separable denoisers 
within AMP for non-i.i.d. signals will track SE.
This confidence motivates us to apply a universal denoiser 
within AMP for CS reconstruction of stationary ergodic signals with
unknown input statistics.
Indeed, the numerical results in Section~\ref{sec:numerical} show that AMP with a universal denoiser leads to a promising universal CS recovery algorithm.

%---------------------------
\section{Universal denoising}
\label{sec:SW}
%--------------------------
%--------------------------------
\subsection{Background}
\label{subsec:SW_old}
%-------------------------------
Sivaramakrishnan and Weissman~\cite{SW_Context2009} propose to quantize the noisy symbols to generate quantized contexts that are used to partition the unquantized symbols into subsequences. That is, given the noisy observations ${\bf q}\in\mathbb{R}^N$, define the context of $q_j$ as
${\bf c}_j=[{\bf q}_{j-k}^{j-1};{\bf q}_{j+1}^{j+k}]$ 
for $j=1+k,...,N-k$, where ${\bf c}_j\in\mathbb{R}^{2k}$, 
and $[ \alpha;\beta ]$ denotes the concatenation 
of the sequences $\alpha$ and $\beta$. 
Vector quantization can be applied to the context set $\mathcal{C}=\lbrace {\bf c}_j:j=1+k,...,N-k \rbrace$, and each ${\bf c}_j$ is assigned a label $l_j\in\lbrace  1,...,L\rbrace
$ that represents the cluster that ${\bf c}_j$ belongs to. Finally, the $L$ subsequences that consist of symbols from ${\bf q}$ with the same label are obtained by taking ${\bf q}_l=\lbrace q_j:l_j=l \rbrace$, for $l=1,...,L$.

The symbols in each subsequence  ${\bf q}_l$ are regarded as approximately
conditionally independent given the common quantized 
contexts~\cite{SW_Context2009}. The rationale underlying this concept is that a sliding-window denoiser uses information from the contexts to estimate the current symbol, and symbols with similar contexts in the noisy output
of the scalar channel have similar contexts in the original signal.
Therefore, symbols with similar contexts can be grouped together 
and denoised using the same denoiser.

Sivaramakrishnan and Weissman~\cite{SW_Context2009} estimate the pdf of ${\bf x}_l$, which is the clean subsequence corresponding to ${\bf q}_l$, by first obtaining an estimate of $p_l(q)$, which is the empirical pdf of ${\bf q}_l$, via quantization followed by kernel density estimation, 
and then quantizing $p_l(x)$ in both the support domain and the probability domain to find a $\widehat{p}_l(x)$ that matches $q_l$ well.
Once $\widehat{p}_l(x)$ is obtained, 
the conditional expectation of the symbols in the $l$-th subsequence can be calculated (\ref{eq:Bayes_scalar}). 

For any well-defined error metric, 
Sivaramakrishnan and Weissman~\cite{SW_Context2009} have proved
for stationary ergodic signals with bounded components that 
their universal denoiser asymptotically achieves the optimal estimation error among all sliding-window denoising schemes despite not knowing the prior for the signal. When the error metric is square error, the optimal error is the MMSE.

%--------------------------------
\subsection{Extension to unbounded signals}
\label{subsec:SW_new}
%-------------------------------
Sivaramakrishnan and Weissman~\cite{SW_Context2009} have shown that one can denoise a stationary ergodic signal by 
({\em i}) grouping together symbols with similar contexts; and ({\em ii}) applying a separable denoiser to each group. 
Such a scheme is optimal in the limit of large signal dimension $N$. However, their denoiser assumes a bounded input, 
which might make it inapplicable to some real-world settings.

We modify step ({\em i}), while employing an entirely different approach for step ({\em ii}) to construct our proposed universal denoiser, which does not require the boundedness condition. 

{\bf Step ({\em i}):} The context set $\mathcal{C}$ is acquired in the same way as described in Section~\ref{subsec:SW_old}, while we add weights to the contexts before clustering. That is, for each ${\bf c}_j\in\mathcal{C}$ of length $2k$, the weighted context is defined as 
\begin{equation}
\label{eq:weighting}
{\bf c}_j'={\bf c}_j\odot{\bf w}, 
\end{equation}
where $\odot$ denotes a point-wise product, and the weights take values, 
\begin{equation}
\label{eq:weight}
w_{k_i} =
\begin{cases}
e^{-\beta (k-k_i)},\quad k_i=1,..,k\\
e^{-\beta (k_i-k-1)},\quad k_i=k+1,...,2k
\end{cases},
\end{equation}
for some $\beta\in (0,1)$,
so that the symbols in ${\bf c}_j$ that are closer in index to $q_j$ have larger weights than the ones that are located farther away. 
The exponential decay rate $\beta$ is made adaptive to the noise level:
\begin{equation}
\label{eq:weight_exp}
\beta = b_1\log_{10}(\sigma_v^2/(\|{\bf q}\|_2^2/N-\sigma_v^2))+b_2,
\end{equation}
where $b_1$ and $b_2$ can be determined numerically.

The weighted context set $\mathcal{C}'=\lbrace {\bf c}'_j:j=1+k,...,N-k \rbrace$ is then sent to a $k$-means algorithm~\cite{MacQueen1967kmeans}, and ${\bf q}_l$'s are obtained according to the labels determined via clustering. 
A post-processing step is added to ensure that the empirical 
pdf of ${\bf q}_l$ is learned from no less than $T$ symbols. That is, if the size of ${\bf q}_l$, which is denoted by $B$, is less than $T$, then $T-B$ symbols in other clusters whose contexts are closest to the centroid of the current cluster are included to estimate the empirical pdf of ${\bf q}_l$, while after the pdf is learned, the extra symbols are removed, and only ${\bf q}_l$ is denoised with the currently learned pdf.
The rationale for requiring at least $T$ symbols is that
using a small amount of data to estimate the pdf may result in
a poor estimate.

{\bf Step ({\em ii}):} In order to overcome the limitation of boundedness, we fit $p(q)$ with a GM model. In the case of Gaussian noise channel, we can then estimate $p(x)$ by subtracting the noise variance from each Gaussian component in $\widehat{p}(q)$. Details are provided in Section~\ref{sec:FJ}.

%---------------------------
\section{i.i.d. denoising via Gaussian mixture fitting}
\label{sec:FJ}
%--------------------------
%--------------------------------
\subsection{Background}
\label{subsec:FJ_old}
%-------------------------------
The pdf of a GM has the form:
\begin{equation}
p(x) = \sum_{s=1}^{S}\alpha_s\mathcal{N}(x;\mu_s,\sigma_s^2),\label{eq:GM}
\end{equation}
where $S$ is the number of Gaussian components, and $\sum_{s=1}^S\alpha_s=1$, so that $p(x)$ is a proper pdf. 

Figueiredo and Jain~\cite{FigueiredoJain2002} propose to fit a GM model for a given data sequence by starting with some arbitrarily large $S$, and inferring the structure of the mixture by letting the mixing probabilities $\alpha_s$ of some components be zero. This approach resembles the concept underlying the minimum message length (MML) criterion that selects the best overall model from the entire model space, which differs from
model class selection based on the best model within each class.\footnote{All models with the same number of components belong to one model class, and different models within a model class have different parameters for each component.} A component-wise EM algorithm that updates $(\alpha_s,\mu_s,\sigma_s^2)$ sequentially in $s$ is used to implement the MML-based approach.
The main feature of the algorithm is that if $\alpha_s$ is estimated as 0, then the $s$-th component is immediately removed, and all the estimates in the 
expectation step are recomputed before moving to the 
maximization step that learns the parameters of the $(s+1)$-th component.

%--------------------------------
\subsection{Extension to denoising}
\label{subsec:FJ_new}
%-------------------------------
Consider the scalar channel denoising problem defined in (\ref{eq:denoising})
with an i.i.d. input. We propose to estimate ${\bf x}$ from its Gaussian noise corrupted observations ${\bf q}$ by posing a GM prior on ${\bf x}$, and learning the parameters of the GM model with a modified version of the algorithm proposed by Figueiredo and Jain~\cite{FigueiredoJain2002}.

{\bf Initialization of EM:}
The EM algorithm must be initialized for each parameter, $\lbrace \alpha_s,\mu_s,\sigma_s^2 \rbrace$, $s=1,...,S$. 
One may choose to initialize the Gaussian components with equal mixing probabilities and equal variances, and the initial value of the means are 
randomly sampled from the input data sequence~\cite{FigueiredoJain2002}.
However, in CS recovery problems, the input signal is often sparse, 
and it becomes difficult to correct the initial value if the initialized
values are far from the truth. 
To see why a poor initialization might be problematic,
consider the following scenario: a sparse binary signal that contains a few ones and is corrupted by Gaussian noise is sent to the algorithm. If the initialization levels of the $\mu_s$'s are all around zero, then the algorithm is likely to fit a Gaussian component with near-zero mean and large variance rather than two narrow Gaussian components, one of which has mean close to zero while the other has mean close to one. 

To address this issue, we modify the initialization to examine the maximal distance between each symbol of the input data sequence and the current initialization of the $\mu_s$'s. If the distance is greater than $0.1\sigma_{\text{init}}$, then we add a Gaussian component whose mean is initialized as the value of the symbol being examined, where 
$\sigma_{\text{init}}^2$ is the initialization level of the variance. We found in our simulations that the modified initialization improves the accuracy of the density estimation, and speeds up the convergence of the EM algorithm; the details of the simulation are omitted for brevity. 

{\bf Leverage side-information about the noise:}
Because we know that each Gaussian component in $\widehat{p}(q)$ should 
have variance no less than the noise variance $\sigma_v^2$, during 
the parameter learning process, if a component has variance that is
significantly less than $\sigma_v^2$, we assume that this
low-variance component is spurious, 
and set the corresponding $\alpha_s$ to zero. 
However, if the variance of the component is only slightly less 
than the noise variance $\sigma_v^2$, then we allow the algorithm 
to keep tracking this component,
and set the component variance to be equal to $\sigma_v^2$
at the end of the parameter learning process. 
That said, when subtracting the noise variance $\sigma_v^2$ from the Gaussian components of $\widehat{p}(q)$ to obtain the components of 
$\widehat{p}(x)$, we could have components with zero-valued variance, which leads to deltas in $\widehat{p}(x)$. 

{\bf Denoising:} 
Once the parameters in (\ref{eq:GM}) are estimated, 
we define a denoiser for i.i.d. signals as conditional expectation:
\begin{align}
\eta_{\text{iid}}(q) &=\mathbb{E}[x|q]\nonumber\\
&=\frac{\int xp(q|x)\widehat{p}(x)\text{d}x}{\widehat{p}(q)}\nonumber\\
&=\frac{\sum_{s=1}^S\alpha_s\mathcal{N}(q;\mu_s,\sigma_s^2+\sigma_v^2)(\frac{\sigma_s^2}{\sigma_s^2+\sigma_v^2}(q-\mu_s)+\mu_s)}{\sum_{s=1}^S\alpha_s\mathcal{N}(q;\mu_s,\sigma_s^2+\sigma_v^2)}.\label{eq:FJ_denoiser}
\end{align}

%--------------------------------
\section{Proposed Universal CS recovery algorithm}
\label{sec:proposed_algo}
%-------------------------------
Combining the three components that have been discussed in 
Sections~\ref{sec:AMP}--\ref{sec:FJ},
we are now ready to introduce our proposed universal CS recovery algorithm 
AMP-UD.

Consider a linear inverse problem~(\ref{eq:matrix_channel}), 
where the input signal ${\bf x}$ is generated by a stationary ergodic source.\footnote{For some measurement matrices such as zero mean Gaussian, 
AMP tends to follow SE. For other matrices, convergence of AMP might be problematic.} To estimate ${\bf x}$ from ${\bf y}$ given ${\bf A}$, 
we apply AMP as defined in~(\ref{eq:AMPiter1}) and (\ref{eq:AMPiter2}). In each iteration, 
observations corrupted by AWGN
${\bf q}^t={\bf x}^t+{\bf A}^T{\bf r}^t$ are obtained. We utilize the 
$k$-means clustering algorithm~\cite{MacQueen1967kmeans}
to perform vector quantization over the weighted contexts~(\ref{eq:weighting}). 
Once we obtain $L$ i.i.d. subsequences $\lbrace{\bf q}_1,...,{\bf q}_L\rbrace$, we fit the empirical pdf of each ${\bf q}_l$ with a GM model using the approach described in Section~\ref{sec:FJ}, and denoise the corresponding subsequence via (\ref{eq:FJ_denoiser}).

To obtain the Onsager correction term in (\ref{eq:AMPiter2}), we need to calculate the derivative of $\eta_{\text{iid}}$ (\ref{eq:FJ_denoiser}). Denoting
\begin{align*}
f(q) &= \sum_{s=1}^S\alpha_s\mathcal{N}(q;\mu_s,\sigma_s^2+\sigma_v^2)(\frac{\sigma_s^2}{\sigma_s^2+\sigma_v^2}(q-\mu_s)+\mu_s),\\
g(q) &= \sum_{s=1}^S\alpha_s\mathcal{N}(q;\mu_s,\sigma_s^2+\sigma_v^2),
\end{align*}
we have that
\begin{align*}
f'(q) &= \sum_{s=1}^S\alpha_s\mathcal{N}(q;\mu_s,\sigma_s^2+\sigma_v^2)\\
&\cdot\left(\frac{\sigma_s^2+\mu_s^2-q\mu_s}{\sigma_s^2+\sigma_v^2}-\left(\frac{\sigma_s(q-\mu_s)}{\sigma_s^2+\sigma_v^2}\right)^2\right),\\
g'(q) &= \sum_{s=1}^S\alpha_s\mathcal{N}(q;\mu_s,\sigma_s^2+\sigma_v^2)\left(- \frac{q-\mu_s}{\sigma_s^2+\sigma_v^2}\right).
\end{align*}
Therefore,
\begin{equation}
\label{eq:denoiser_FJ_deriv}
\eta_{\text{iid}}^{'}(q)=\frac{\text{d}}{\text{d}q}\eta_{\text{iid}}(q)=\frac{f'(q)g(q)-f(q)g'(q)}{(g(q))^2}.
\end{equation}

\begin{CONJ}
Consider a linear inverse problem~(\ref{eq:matrix_channel}) where we want to estimate the input signal ${\bf x}$ from its noisy measurements ${\bf y}$ and the measurement matrix ${\bf A}$. Under some technical conditions, the AMP-UD algorithm, where the proposed universal denoiser based on context quantization and GM-based i.i.d. subsequence denoising~(\ref{eq:FJ_denoiser}), is applied within AMP iterations~(\ref{eq:AMPiter1},~\ref{eq:AMPiter2}), achieves the MMSE.
\end{CONJ}

%------------------
\section{Numerical results}
\label{sec:numerical}
%------------------
We implement AMP-UD in Matlab 
(we are in the process of posting our code online)
on a Dell OPTIPLEX 9010 running an Intel(R) 
$\text{Core}^{\text{TM}}$ i7-3770 with 16GB RAM, and test it utilizing different types of signals at various measurement rates and SNR levels, where SNR is defined as $\text{SNR}=10\log_{10}[(N\mathbb{E}[x^2])/(M\sigma_z^2)]$.  The input signal length $N$ is 10,000.
The context quantization is implemented via the $k$-means algorithm~\cite{MacQueen1967kmeans}, and the initial number of clusters is set to be 10. That is, $L$ is initialized as 10, and may become smaller if empty clusters occur. The lower bound $T$ on the number of symbols required to learn the GM parameters is $256$. 
The context size $2k$ is chosen to be 12, and the contexts are weighted according to (\ref{eq:weight}) and (\ref{eq:weight_exp}). In order to avoid possible divergence of AMP-UD, we employ a damping technique~\cite{Rangan2014ISIT} to slow down the evolution. Specifically, damping is an extra step in the AMP iteration~(\ref{eq:AMPiter2}), instead of updating the value of ${\bf x}^{t+1}$ by the output of the denoiser $\eta^t({\bf A}^T{\bf r}^t+{\bf x}^t)$, a weighted sum of $\eta^t({\bf A}^T{\bf r}^t+{\bf x}^t)$ and ${\bf x}^t$ is taken as follows,
\begin{equation}
\label{eq:damp}
{\bf x}^{t+1}=\lambda\eta^t({\bf A}^T{\bf r}^t+{\bf x}^t)+(1-\lambda){\bf x}^t,
\end{equation} 
for some constant $0<\lambda\le 1$. In our simulation, $\lambda$ is set to be 0.1, and we run 100 AMP iterations.
The recovery performance is evaluated by signal to distortion ratio ($\text{SDR}=10\log_{10}(\mathbb{E}[x^2]/\text{MSE})$), where the MSE is averaged over 50 random draws of ${\bf x}$, ${\bf A}$, and ${\bf z}$. 

We compare the performance of AMP-UD to ({\em i}) 
the universal CS recovery algorithm SLA-MCMC~\cite{ZhuBaronDuarte2014_SLAM}; and
({\em ii}) the Bayesian message passing approaches EM-GM-AMP-MOS~\cite{EMGMTSP} for i.i.d. inputs and turboGAMP~\cite{turboGAMP} for non-i.i.d. inputs. Note that EM-GM-AMP-MOS 
assumes during recovery that the input is i.i.d., 
whereas turboGMAP is designed for non-i.i.d. inputs 
with a known statistical model.
We do not include results for other well-known CS
algorithms such as compressive sensing matching pursuit 
(CoSaMP)~\cite{Cosamp08} or gradient projection for sparse 
reconstruction (GPSR)~\cite{GPSR2007}, because their SDR
performance is consistently weaker than the three algorithms
being compared.

\begin{figure}[t!]
\center
\includegraphics[width=85mm]{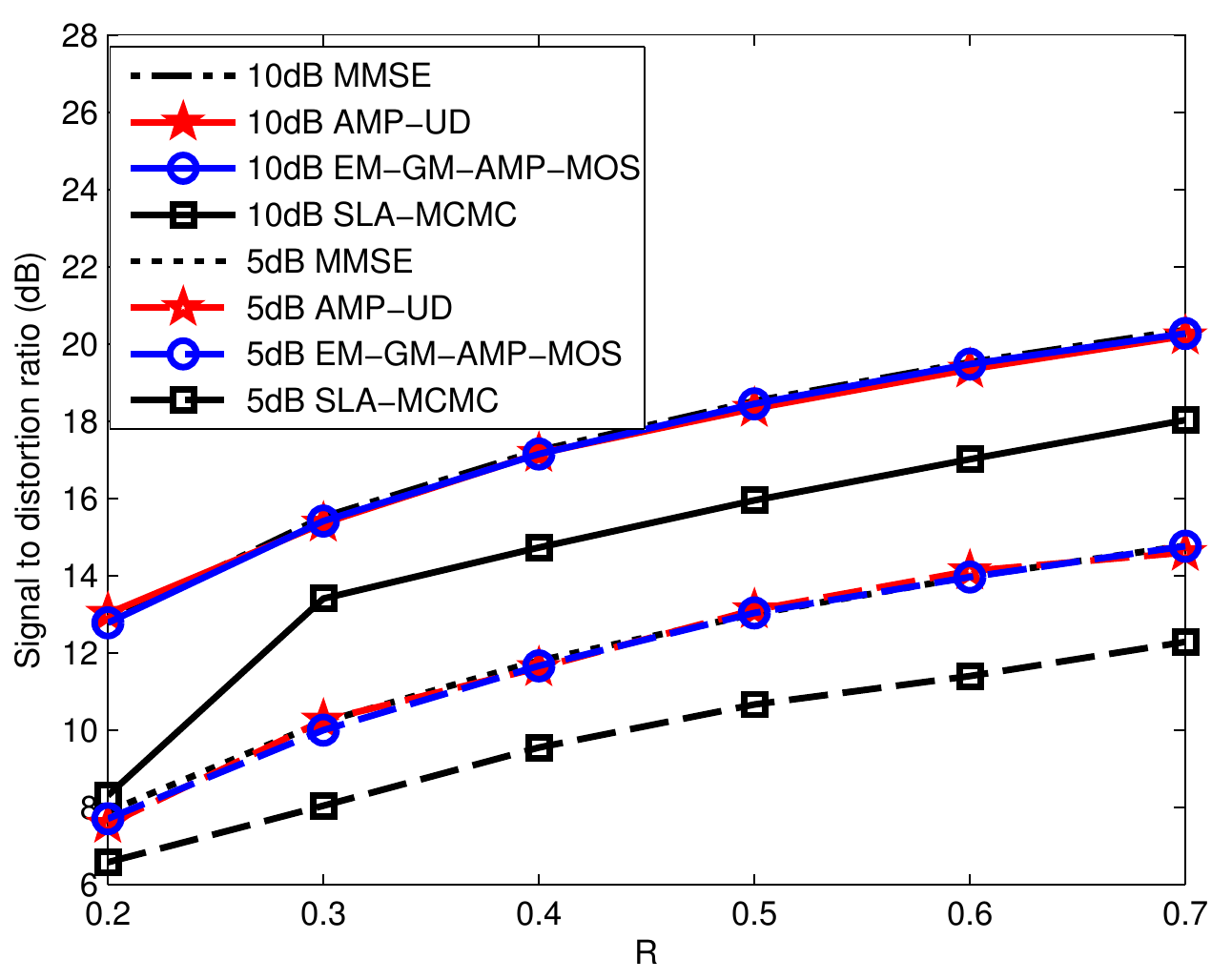}
\caption{AMP-UD, SLA-MCMC, and EM-GM-AMP-MOS reconstruction results for an i.i.d. sparse Laplace signal as a function of measurement rate. ($N=10,000$, SNR = 5 dB or 10 dB.)}
\label{fig:AMP_Laplace}
\vspace*{-5mm}
\end{figure}

{\bf i.i.d. sparse Laplace signal:}
An i.i.d. sparse Laplace signal follows the distribution $p(x)=0.03\mathcal{L}(0,1)+0.97\delta(x)$, where $\mathcal{L}(0,1)$ denotes a Laplacian distribution with mean zero and variance one, and $\delta(\cdot)$ is the delta function~\cite{Papoulis91}. 
It is shown in Fig.~\ref{fig:AMP_Laplace} that AMP-UD and EM-GM-AMP-MOS achieve the MMSE, whereas SLA-MCMC has 
weaker performance, because the MCMC approach is expected to sample 
from the posterior and its MSE is twice the 
MMSE~\cite{DonohoKolmogorov,ZhuBaronDuarte2014_SLAM}.

\begin{figure}[t!]
\center
\includegraphics[width=85mm]{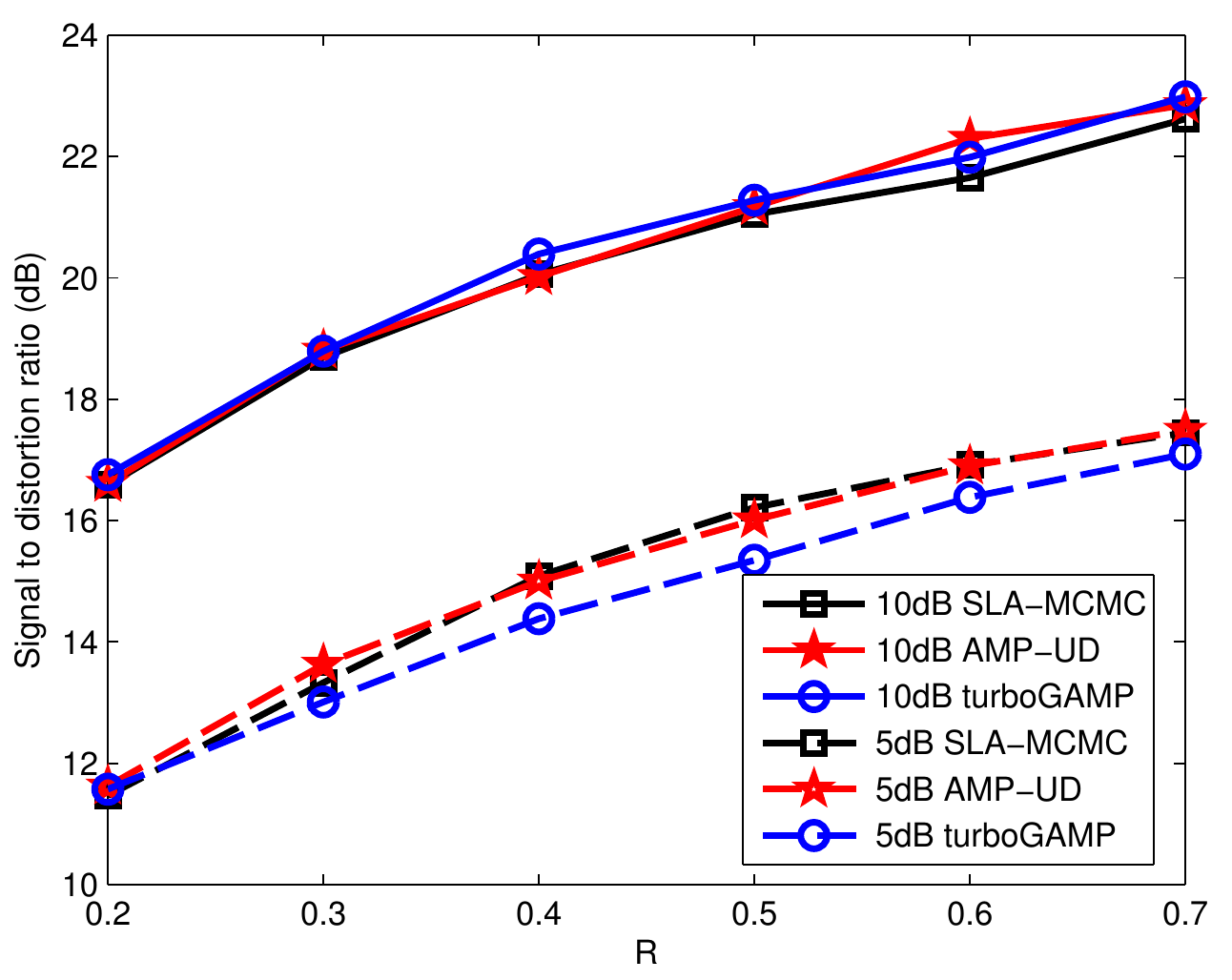}
\caption{AMP-UD, SLA-MCMC, and turboGAMP reconstruction results for a two-state Markov signal with nonzero entries drawn from a uniform distribution $U[0,1]$ as a function of measurement rate. 
($N=10,000$, SNR = 5 dB or 10 dB.)}
\label{fig:AMP_MUnif}
\vspace*{-5mm}
\end{figure}

{\bf Markov-uniform signal:}
Consider the two-state Markov state machine defined in Section~\ref{subsec:SE} with $p_{01}=\frac{3}{970}$ and $p_{10}=0.10$. A Markov-uniform signal (MUnif for short) follows a uniform distribution $U[0,1]$ at 
the nonzero state $s_1$. 
These parameters lead to 3\% nonzero entries in a MUnif signal on average.
It is shown in Fig.~\ref{fig:AMP_MUnif} that at low SNR, AMP-UD and SLA-MCMC have similar SDR performance and are both better than turboGAMP despite not knowing the Markovian structure of the signal. At high SNR, the three algorithms are comparable in SDR.

\begin{figure}[t!]
\center
\includegraphics[width=85mm]{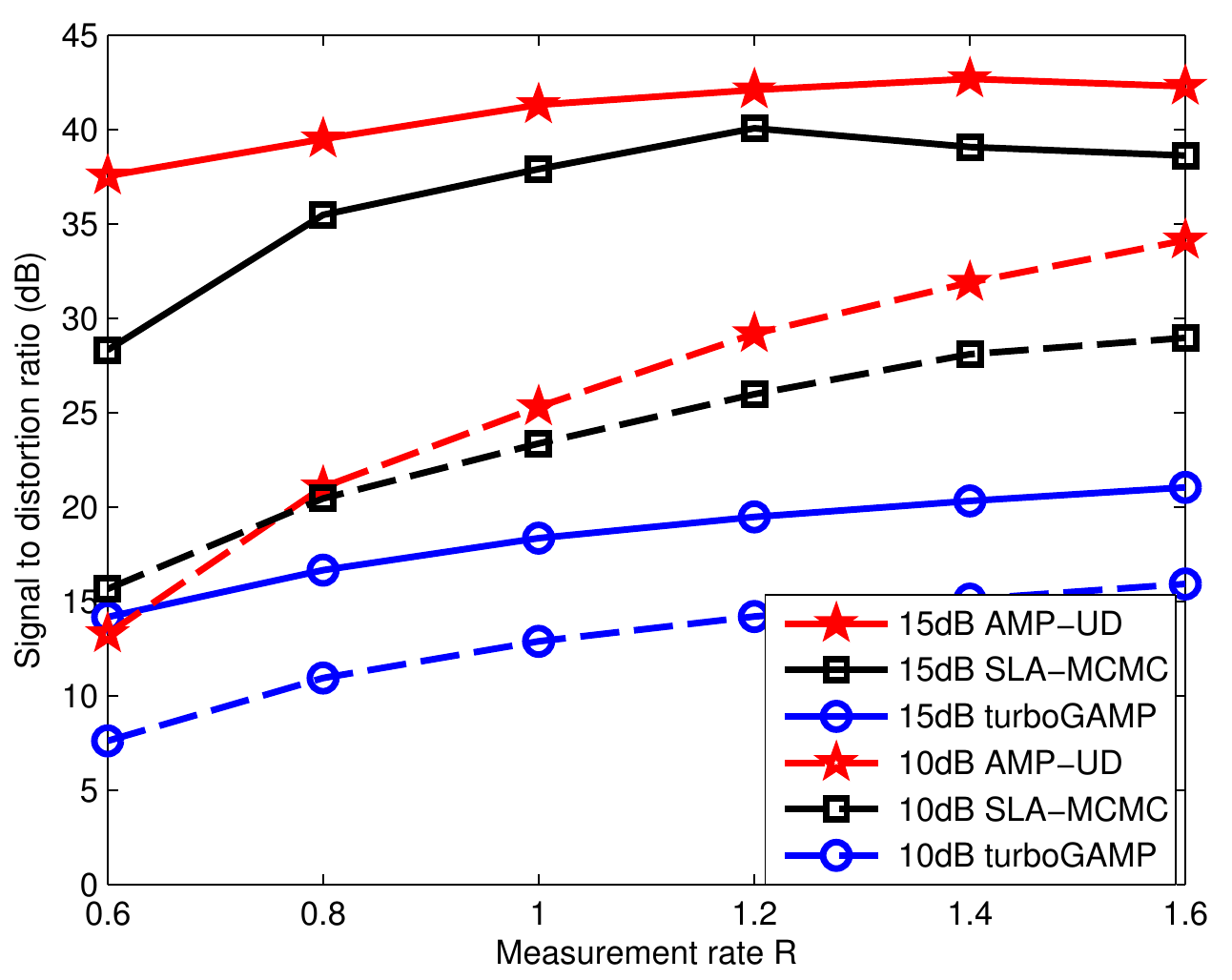}
\caption{AMP-UD, SLA-MCMC, and turboGAMP reconstruction results for a dense two-state Markov signal with nonzero entries drawn from a Rademacher ($\pm 1$) distribution as a function of measurement rate. 
($N=10,000$, SNR = 10 dB or 15 dB.)}
\label{fig:AMP_MRad}
\vspace*{-5mm}
\end{figure}

{\bf Dense Markov Rademacher signal:}
Consider the two-state Markov state machine defined in 
Section~\ref{subsec:SE} with $p_{01}=\frac{3}{70}$ and $p_{10}=0.10$. 
A dense Markov Rademacher signal (MRad for short) 
takes values from $\lbrace -1,+1  \rbrace$ with equal probability 
at $s_1$. These parameters lead to 30\% nonzero entries in an 
MRad signal on average. Because the MRad signal is dense 
(non-sparse), we must measure it with somewhat larger 
measurement rates and SNRs than before.
It is shown in Fig.~\ref{fig:AMP_MRad} that AMP-UD and SLA-MCMC achieve better overall performance than turboGAMP despite not knowing the Markovian signal structure. AMP-UD outperforms SLA-MCMC except for the lowest tested measurement rate at low SNR. That said, further tests suggest
that AMP-UD can be improved in this configuration.

{\bf Runtime:} The runtime of AMP-UD for MUnif and MRad is typically under 10 minutes, but somewhat more for signals such as sparse Laplace that require a large number of Gaussian components to be fit. For comparison, the runtime of SLA-MCMC is typically an hour, whereas typical runtimes of EM-GM-AMP-MOS and turboGAMP are 30 minutes.

%------------------
\section{Conclusion}
\label{sec:conclusion}
%------------------
In this paper, we introduced a universal 
CS recovery algorithm AMP-UD that applies our proposed
universal denoiser (UD) within approximate message passing (AMP).
AMP-UD is designed to reconstruct stationary ergodic inputs from
noisy linear measurements. 
Although AMP-UD is universal and does not know the input statistics,
it achieves favorable signal to distortion ratios  compared to existing algorithms,
and its runtime is typically faster.

The algorithm combines three existing schemes:
({\em i}) AMP~\cite{DMM2009}; 
({\em ii}) universal denoising~\cite{SW_Context2009}; and
({\em iii}) a density estimation approach based on Gaussian mixture fitting~\cite{FigueiredoJain2002}.
In addition to the algorithmic framework, we provided three contributions. 
First, we provided numerical results showing that SE holds for non-separable Bayesian sliding-window denoisers.
Second, we designed a universal denoiser that does not require the input signal to be bounded.
Third, we modified the GM learning algorithm, and extended it to an i.i.d. denoiser.

There are numerous directions for future work.
First, can we provide theoretical guarantees that our denoiser asymptotically  achieves the MMSE for unknown
stationary ergodic signals? Such guarantees were proven by Sivaramkrishnan and Weissman~\cite{SW_Context2009}, but we 
have extended their approach to unbounded signals.
Second, it is not clear whether our Gaussian mixture fit can deal with outliers, and we aim to develop a robust
density estimation scheme.
Third, our current algorithm was designed to minimize the square error, and the denoiser could be modified to minimize
other error metrics~\cite{Tan2014}.
Finally, AMP-UD was designed to reconstruct one-dimensional signals. In order to support
applications that process multi-dimensional  signals such as images, it might be instructive to employ universal image denoisers within AMP.
\section*{Acknowledgements}
We thank Mario Figueiredo and Tsachy Weissman for informative discussions.

\ifCLASSOPTIONcaptionsoff
\newpage
\fi
\bibliographystyle{IEEEtran}
\bibliography{IEEEabrv,cites}
\end{document}